# Anomalous Light Scattering by Topological $\mathcal{PT}$-symmetric Particle Arrays

C. W. Ling,[1] Ka Hei Choi,[1] T. C. Mok,[1] Zhao-Qing Zhang,[2] and Kin Hung Fung[1, *]

[1]*Department of Applied Physics, The Hong Kong Polytechnic University, Hong Kong, China*
[2]*Department of Physics, The Hong Kong University of Science and Technology, Hong Kong, China*
(Dated: June 24, 2016)

Robust topological edge modes may evolve into complex-frequency modes when a physical system becomes non-Hermitian. We show that, while having negligible forward optical extinction cross section, a conjugate pair of such complex topological edge modes in a non-Hermitian $\mathcal{PT}$-symmetric system can give rise to an anomalous sideway scattering when they are simultaneously excited by a plane wave. We propose a realization of such scattering state in a linear array of subwavelength resonators coated with gain media. The prediction is based on an analytical two-band model and verified by rigorous numerical simulation using multiple-multipole scattering theory. The result suggests an extreme situation where leakage of classical information is unnoticeable to the transmitter and the receiver when such a $\mathcal{PT}$-symmetric unit is inserted into the communication channel.

## INTRODUCTION

Parity-time ($\mathcal{PT}$)-symmetric quantum mechanics[1] has opened up a new direction in searching for unconventional states of matter. Non-Hermitian $\mathcal{PT}$-symmetric systems have been a subject of intense studies because they exhibit unusual properties, such as $\mathcal{PT}$-phase transitions[2]. While realization in quantum systems could be in doubt, several unconventional $\mathcal{PT}$-symmetry related phenomena, such as transition from real-frequency modes to conjugate pair of complex-frequency modes, have been realized in classical photonic systems[2–6]. Recently, many efforts have been put on extending topological band theory[7,8] to non-Hermitian $\mathcal{PT}$-symmetric systems. For example, there have been different theoretical approaches to generalize topological invariants using bi-orthonormal basis[9–12], redefining the inner product[13], or using the global Berry phase[14]. Topological transition in the bulk of non-Hermitian system has also been realized[15]. It has been proposed that eigenstates of these non-Hermitian system associated with exceptional points could lead to new physics and applications, such as realization of Majorana zero modes[16,17] and single mode lasers[18].

Topological edge modes may also evolve into complex-frequency modes when the system becomes non-Hermitian[19–21]. The physical consequence of such complex topological edge modes, which decay in both space and time, is obscure so far. In this paper, we suggest a way to realize complex-frequency topological edge modes in $\mathcal{PT}$-symmetric photonic systems. We show that a conjugate pair of topological complex-frequency edge modes in a $\mathcal{PT}$-symmetric photonic systems can be realized through observing an anomalous sideway scattering by an array of subwavelength resonators coated with gain media. This suggests an extreme situation where leakage of classical information is unnoticeable to the transmitter and the receiver when such a $\mathcal{PT}$-symmetric unit is inserted to the communication channel.

## RESULTS

### 2-band $\mathcal{PT}$-symmetric model

We begin with the topological description of a non-Hermitian $\mathcal{PT}$-symmetric periodic system. It can be shown that, even if the edge mode frequencies become complex in non-Hermitian $\mathcal{PT}$- symmetric systems, the Zak geometrical phase of a bulk band is still quantized as 0 or $\pi$ when all bulk modes are in the unbroken $\mathcal{PT}$-symmetric regime. To provide a complete description and proof, we start by considering a two-band, non-Hermitian $\mathcal{PT}$-symmetric model which can be used to describe an array of subwavelength photonic resonators coated with gain media. The $\mathcal{PT}$-symmetric eigenvalue problem is written in the generic form as[14,22]

$$\mathbf{H}_k \mathbf{u} = E_k \mathbf{u}, \quad (1)$$

where

$$\mathbf{H}_k = \begin{bmatrix} ih_z(k) & h_x(k) - ih_y(k) \\ h_x(k) + ih_y(k) & -ih_z(k) \end{bmatrix}, \quad (2)$$

$E_k$ is the eigenvalue, $\mathbf{u}$ is the right eigenvector, $k$ is the Bloch wave vector, and $d$ is the lattice constant.

It should be noted that $h_x(k)$, $h_y(k)$, and $h_z(k)$ are periodic real-valued functions with period of $2\pi/d$ in $k$, and they form a $\vec{h}$-vector space as shown in Figure 1(a). The subscripts $x$, $y$, and $z$ refer to the directions in the $\vec{h}$-vector space (not the real space), while the expressions are determined by the actual approximation. Each system is represented as a closed loop $\mathcal{C}$ in Fig. 1(a) as the parameter $k$ varies from $-\pi/d$ to $\pi/d$. This closed loop is associated with a winding number which determines the Zak geometric phase of a band.

The Zak phase of each band ($+$ or $-$) is usually ex-

*[*] khfung@polyu.edu.hk

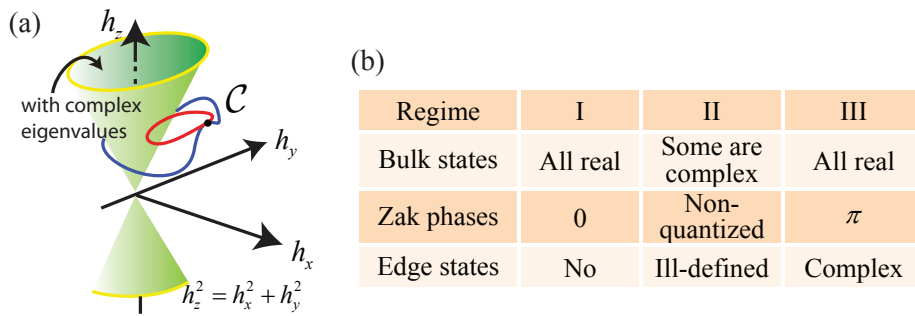

FIG. 1. (Color online) Classification of topological bands and edge modes of a $\mathcal{PT}$-symmetric system described by Eq. (1). (a) Two types of paths in parameter space. The blue and the red path have winding number 1 and 0, respectively. (b) Table summarizing the three regimes according to the bulk and the edge state frequencies, including the Zak phase. In regimes I and III, the bulk states have real eigenvalues and the entire bulk dispersion relation is in the unbroken $\mathcal{PT}$-symmetric phase. In regime II, some bulk states are associated with complex eigenvalues and thus part of the bulk states are in the broken $\mathcal{PT}$-symmetric phase.

pressed as[23–27]

$$\gamma_\pm = i \int_{-\frac{\pi}{d}}^{\frac{\pi}{d}} \mathbf{u}_\pm^*(k) \cdot \frac{d}{dk} \mathbf{u}_\pm(k) dk. \qquad (3)$$

The analytical evaluation of Zak phase is given in the Methods. When $E_{k\pm}$ are real for all $k \in [-\pi/d, \pi/d]$ (i.e., all bulk states in the unbroken $\mathcal{PT}$-symmetric regime), the whole loop $\mathcal{C}$ is outside the kissing cones $h_x^2 + h_y^2 > h_z^2$ [see Fig. 1(a)]. Under this condition, we have quantized Zak phase

$$\gamma_\pm = w\pi, \qquad (4)$$

where $w$ is the (integer) winding number of $\vec{h}(k)$ about the $h_z$ axis as $k$ varies. It can also be shown that all even (odd) integer values of $w$ are equivalent and Zak phase is thus usually quantized as zero or $\pi$. Since there exist different approaches in defining the Zak phase, we also provide a comparison to the results in bi-orthonormal basis (see Supplementary Information).

Zak phase can be used to classify the bulk band topology due to gauge invariance[28]. More importantly, its value can help us predict the existence of the edge modes when all bulk states are in the unbroken $\mathcal{PT}$-symmetric phase. For a large but finite $\mathcal{PT}$-symmetric system, the weakly coupled edge modes also guarantee the existence in form of conjugate pair. In this case, the system can be classified (according to the bulk states) as three regimes summarized in Fig. 1(b). In this paper, we focus on the complex edge modes in regime III which can be simultaneously excited by a plane wave. In general, when $\mathcal{P}$ symmetry is broken, a linear combination of the edge modes can be excited so that the symmetry of the scattered waves may not follow the spatial symmetry of the plane wave. In the next section, we show that when parity ($\mathcal{P}$) and time-reversal ($\mathcal{T}$) symmetries are broken but $\mathcal{PT}$ symmetry is not broken, the spatial symmetry of the scattered waves can be nearly opposite to that of the excitation plane wave, which finally gives rise to a negligible forward optical extinction but non-zero sideway scattering.

### $\mathcal{PT}$-symmetric plasmonic particle array

We consider a spatially periodic plasmonic structure in which the electromagnetic modes are represented by the eigenvalue problem equivalent to Eq. (1) and takes the form similar to Bergman's representation[29–31]:

$$\mathbf{A}_k \mathbf{p}_k = \epsilon_1 \mathbf{p}_k, \qquad (5)$$

where $k$, $\mathbf{p}_k$, and $\epsilon_1$ are the Bloch wave vector, eigenvector, and eigenvalue respectively. The problem corresponds to a dimer array formed by two types of coated spherical plasmonic nanoparticles with alternative separations, $s$ and $d-s$, as shown in Fig. 2(a). The coating shells are non-dispersive dielectrics, and are gain-loss-balanced[32].

Eq. (5) comes from the coupled dipole equation[23,33–35], which is obtained by approximating each nanoparticle as a point dipole scatter, using the Bloch's theorem, and taking the quasistatic approximation (see Methods). The dipole moments on nanoparticle $A$ and $B$ are denoted by the vector $\mathbf{p}_k = (p_{k;A}, p_{k;B})^\mathrm{T}$, while the matrix $\mathbf{A}_k$ is related to dipole couplings and the polarizabilties. Eigenvalue $\epsilon_1$ is also the dielectric function of the plasmonic core, and is mapped to the plasmon frequency $\omega$ via the Drude model[23,33–35], given by

$$\omega/\omega_p = \frac{1}{2}\left(\sqrt{\frac{4}{1-\epsilon_1} - \frac{1}{\tau^2}} - \frac{1}{\tau}i\right), \qquad (6)$$

in which $\omega_p$ and $1/\tau$ are the plasma frequency and electron collision frequency for the plasmonic core.

The finite plasmonic particle array has no $\mathcal{P}$ and $\mathcal{T}$ symmetries but $\mathcal{PT}$-symmetry when the plasmonic cores

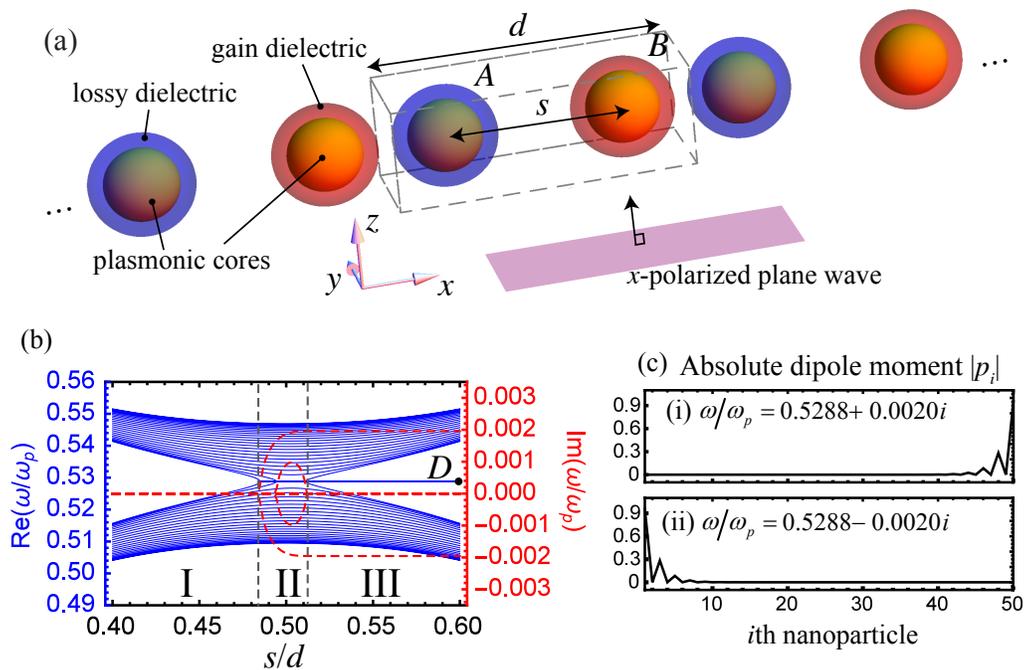

FIG. 2. (Color online) Eigenmodes of the $\mathcal{PT}$-symmetric plasmonic particle array calculated based on coupled dipole method. (a) Schematic of the $\mathcal{PT}$-symmetric plasmonic particle array, which is an array of alternatively coated nanoparticles. The coated nanoparticles have inner radius $a$ and outer radius $b$. A unit cell contains nanoparticle $A$ and $B$ with identical sizes while their shells are lossy and gain dielectrics with $\epsilon_3$ and $\epsilon_3^*$ as the dielectric constants (gain-loss-balanced). In (b) and (c), we set $a = 0.125d$, $b = 0.175d$, and $\epsilon_3 = 1.5 + 0.025i$ for the plasmonic particle array. (b) Resonant frequencies $\omega$ of a finite array (25 unit cells) against $s$. The bands are classified into three regimes (I to III). The blue horizontal line in regime III is associated with the dashed lines at $\text{Im}(\omega/\omega_p) = \pm 0.002$, which corresponds to a pair of topological edge modes. (c) Dipole moment patterns (absolute value) of the topological edge modes at point $D$ in (b), where $s = 0.6d$.

are lossless ($1/\tau = 0$). A detailed discussion on symmetry operators is given in the Methods.

### Complex edge modes

The longitudinal mode frequencies[36] of a finite $\mathcal{PT}$-symmetric particle array when $1/\tau = 0$ are shown in Fig. 2(b)[37]. As described by the table in Fig. 1(b), there are three regimes (I to III) in Fig. 2(b). In regime I, two bands can be clearly identified, and all $\omega$ are real. This implies all the eigenvalues $\epsilon_1$ are real, so the entire bulk dispersion relation is in unbroken $\mathcal{PT}$-symmetric phase. Regime III is similar to regime I except that there are two complex topological edge modes supported in the band gap in regime III. The frequencies of these two topological edge modes form a complex conjugate pair, which is associated with the horizontal blue line and the red lines at $\text{Im}(\omega/\omega_p) = \pm 0.002$. The existence of these complex edge modes are due to $\gamma = \pi$, which is classified as the (non-trivial) regime III. The evaluation of Zak phase for different gain/loss parameters are provided in Supplementary Information.

The two edge modes are eigenmodes with simultaneous complex frequency and complex wavevector, meaning that they decay or grow in both space and time. If the array is long enough, we have the following closed form solution (see Methods):

$$p_{n;\sigma}(t) \propto e^{\pm \text{Im}(\omega_0)t} e^{\pm n\kappa d}, \qquad (7)$$

where $\kappa = (3/d)\ln[s/(d-s)]$, $\omega_0$ is the complex frequency of an edge mode, and $p_{n;\sigma}(t)$ represents the time-domain dipole moments in the $n$th unit cell. Both of the edge modes decay spatially, and the mode patterns at point $D$ are shown in Fig. 2(c) for reference.

### Vanishing optical extinction cross section at resonance

Here, we first explain an unusual zero optical extinction in a situation where the topological complex modes are excited. To excite the topological edge modes in regime III, we introduce a $x$-polarized electrical plane wave with magnitude $E_0$ and frequency $\omega$, as shown in Fig. 2(a). The excited frequency-domain dipole moments on the $2N$ nanoparticles $\mathbf{p} = (p_1, p_2, ...p_{2N})^{\text{T}}$ are linearly dependent on the external wave $\mathbf{E}_0 = (E_0, E_0, ..., E_0)^{\text{T}}$

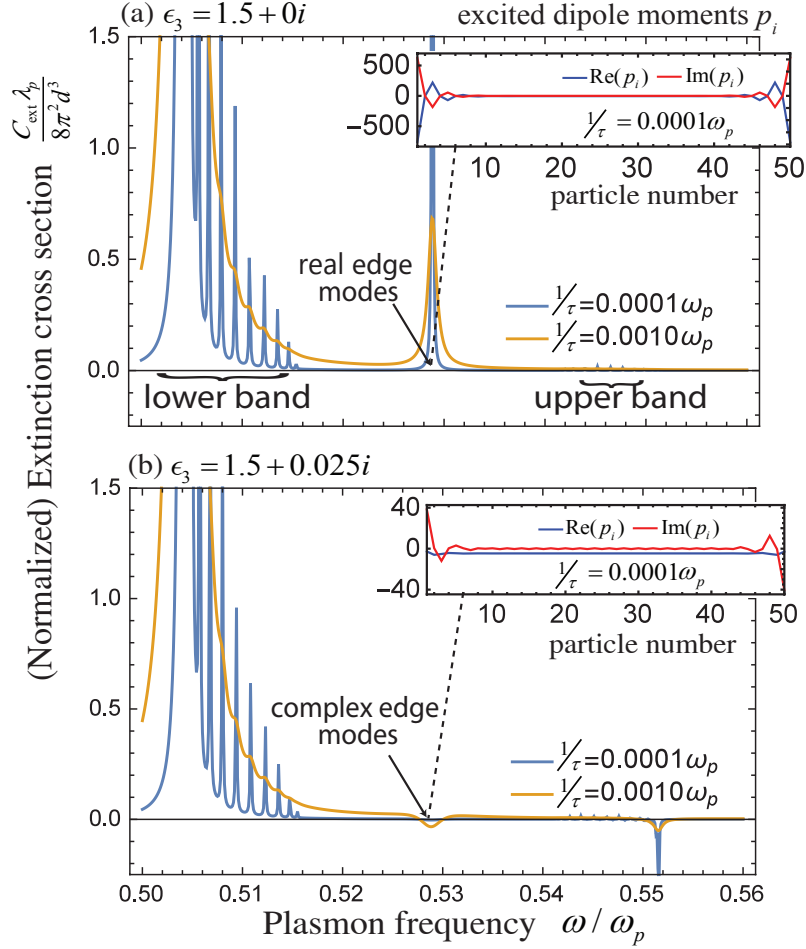

FIG. 3. (Color online) Comparison of optical extinction cross sections by plane waves between (a) a normal particle array and (b) a $\mathcal{PT}$-symmetric particle array with inner separation $s = 0.6d$. $\lambda_p = 2\pi c/\omega_p$. The extinctions dropped at $\omega = 0.529\omega_p$ with small finite Im($\epsilon_3$). The embedded subfigures are the excited dipole moments in each case at $\omega = 0.529\omega_p$. Other parameters are $s = 0.6d$, $a = 0.125d$, $b = 0.175d$, and $N = 25$.

(see Methods):

$$\mathbf{p} = \mathbf{S}(\epsilon_1)\mathbf{E}_0, \quad (8)$$

where $\mathbf{S}(\epsilon_1)$ is a square matrix depending on $\epsilon_1$, and explicit expression is shown in Eq. (21).

In Fig. 3, we show the corresponding extinction cross section calculated in dipole approximation[38,39]

$$C_{\text{ext}} = \frac{k_0}{\epsilon_0 |E_0|^2} \text{Im} \sum_{i=1}^{2N} E_0^* p_i, \quad (9)$$

where $k_0 = \omega/c$, $c$, and $\epsilon_0$ are, respectively, the photon wavenumber, the speed of light, and the permittivity in free space.

In Fig. 3(a), the peak at $\omega = 0.529\omega_p$ is due to the conjugate pair of edge modes, which drops drastically when Im($\epsilon_3$) slightly increases from 0 [Fig. 3(a)] to 0.025 [Fig. 3(b)]. Surprisingly, the two edge modes are strongly excited in Fig. 3(b) (as shown by the normalized dipole moment in the inset) while the corresponding optical extinction cross section is vanishing.

To explain the vanishing $C_{\text{ext}}$ in Fig. 3(b), we define and evaluate the symmetry operators (see Methods for details). The time reverse operator $\mathcal{T}$ turns $p_i$ into its complex conjugate ($\mathcal{T}p_i = p_i^*$). On the other hand, the inversion operator $\mathcal{P}$ flips the direction of $p_i$ and reverse the position order of the nanoparticles ($\mathcal{P}p_i = -p_{2N-i+1}$). Consequently, their combination, transforms a column vector $p_i$ and a general $2N \times 2N$ matrix $A_{i,j}$ in the way that $\mathcal{PT}p_i = -p^*_{2N-i+1}$. Since the matrices $\mathbf{A}$ and $\mathbf{S}(\epsilon_1)$ commute with $\mathcal{PT}$ when $\epsilon_1$ is real (see Methods for a proof), we have $\mathcal{PT}\mathbf{p} = \mathcal{PT}(\mathbf{S}(\epsilon_1)\mathbf{E}_0) = \mathbf{S}(\epsilon_1)\mathcal{PT}\mathbf{E}_0 = -\mathbf{S}(\epsilon_1)\mathbf{E}_0$ and thus

$$p_{2N-i+1} = \sum_{j=1}^{2N} S(\epsilon_1)^*_{i,j} E_0. \quad (10)$$

when the incident field is uniform (i.e., $E_{0i} = E_0$). The result is consistent with the actual dipole moments (ac-cording to Eq. (8)) as shown in the insets of Fig. 3(b) by setting $E_0 = 1$. At the frequencies when only complex eigenmodes are excited (i.e., no divergence of extinction due to real eignemode), we have

$$C_{\text{ext}} = \frac{k_0}{\epsilon_0 |E_0|^2} \text{Im} \sum_{i=1}^{N} E_0^*(p_i + p_{2N-i+1})$$
$$= \frac{k_0}{\epsilon_0} \text{Im} \sum_{i=1}^{N} \sum_{j=1}^{2N} [S(\epsilon_1)_{i,j} + S(\epsilon_1)_{i,j}^*] = 0,$$
(11)

which means that the simultaneously excited edge modes give vanishing $C_{\text{ext}}$ in the perfect $\mathcal{PT}$ situation.

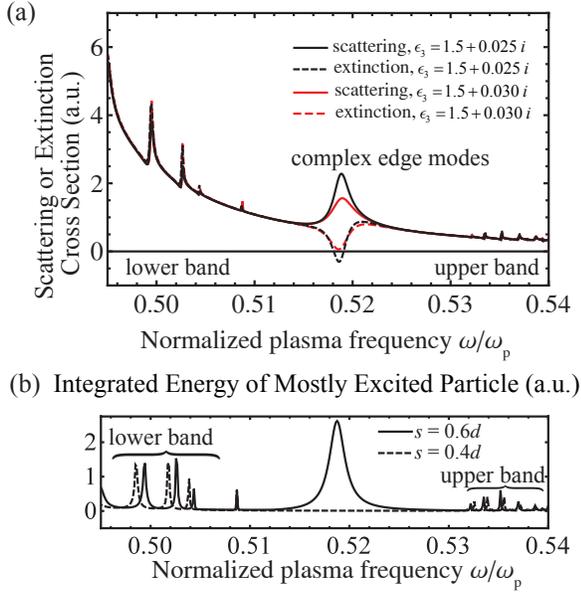

FIG. 4. (Color online) Verification by MST with the contribution from multipoles up to the 5th order. (a) Scattering and extinction cross sections of $\mathcal{PT}$-symmetric particle arrays with $\epsilon_3 = 1.5+0.025i$ and $1.5+0.03i$ ($s = 0.6d$). (b) Integrated field intensity of the mostly excited particle when $\epsilon_3 = 1.5+0.025i$. The dashed peak in (b) at $\omega = 0.5188\omega_p$ indicated that the edge modes are strongly excited simultaneously. Other parameters are $a = 0.125d$, $b = 0.175d$, $d = 100$nm, $N = 16$, and $1/\tau = 0.0001\omega_p$.

It should be noted that when real eigenmodes are excited, the $C_{\text{ext}}$ should normally diverge at the resonance frequencies. In the case of complex eigenmode being excited, the $C_{\text{ext}}$ would be finite. In discussion above, we conclude that extinction cross section is vanishing even if the complex topological edge modes are excited. Due to finite absorption and radiation effects, we found that the net extinction cross section could be below zero, which means that there is net finite amount of energy emitted from the array of particles. As shown in Fig. 3(b), it is clear that an anti-symmetric response of the array is shown, which will contribute to net radiation but the field is orthogonal to the plane wave such that it does not contribute to forward or backward scattering along the propagating direction of incident wave.

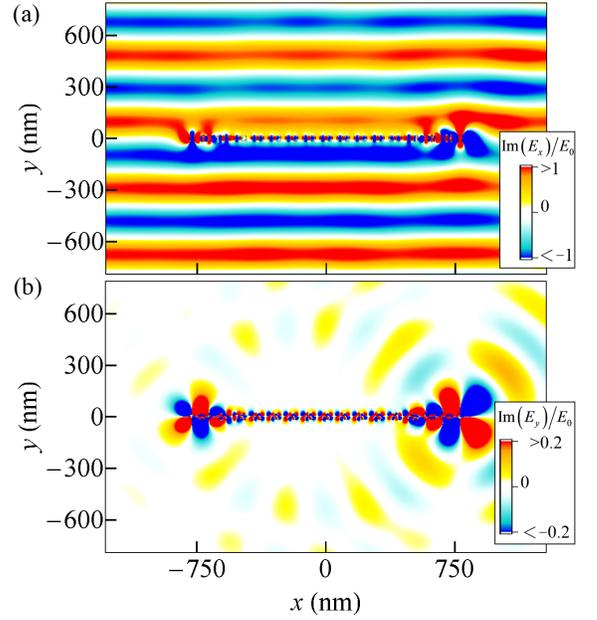

FIG. 5. (Color online) Electric field patterns at $\omega = 0.5188\omega_p$ of the $\mathcal{PT}$-symmetric particle array ($\epsilon_3 = 1.5+0.025i$) demonstrated in Fig. 4(a). Panel (a) and (b) display the $x$ and the $y$ components of the field. The real part of $E_y$ component is not shown for simplicity since it is dominated by the imaginary part. A symmetric distribution of $E_y$ means an anti-symmetric distribution of dipole moments. Due to anti-symmetric resonance response, the array gives strong off-normal scattering but weak distortion to the (forward) transmission and (backward) reflection. Other parameters are $a = 0.125d$, $b = 0.175d$, $d = 100$nm, $N = 16$, and $1/\tau = 0.0001\omega_p$.

**Verifications by multiple scattering theory**

We present the verifications by Multiple Scattering Theory (MST)[40,41] which includes up to the 5th order multipoles as well as dynamic interactions in Figs. 4 and 5. The scattering and extinction cross sections of $\mathcal{PT}$-symmetric particle arrays calculated by MST are shown in Fig. 4. Remarkably, it is shown in Fig. 4(a) that the simultaneously excited edge modes ($\omega \sim 0.52\omega_p$) provide strong scattering cross section but negligible extinction cross section. This is due to the net emission of energy associated with the slightly imbalance between gain and loss since there exists additional radiation loss in realistic situation beyond the quasi-static dipole approximation in previous sections.

In Fig. 4(b), we show the power spectrum absorbed/emitted by the mostly excited particle, which also represents the integrated electric field intensity in each particle. The dashed peak at $\omega = 0.5188\omega_p$ for the case of $s = 0.6d$ is produced by the simultaneously excited edge modes. A comparison with the case of $s = 0.4d$ shows the significant contribution from the edge modes. This provides another piece of evidence that the edge modes are strongly excited even though extinction cross section $C_{\text{ext}}$ is vanishing.

To show the anti-symmetric response of the particle array, we plot the electric field patterns of the $\mathcal{PT}$-symmetric particle array at edge mode frequency in Fig. 5. The $x$- and $y$-components of the electric field are plotted in Figs. 5(a) and (b), respectively. It is observed that the distortion on the $E_x$ component of the electric field is small, which is due to the fact that $E_x$ mostly represents the light transmitted or reflected in the forward or backward directions. Moreover, Fig. 5(b) shows that $E_y$ is close to symmetric with reference to an inversion about the center of the array, which represents a nearly anti-symmetric response. Field patterns of the same particle array without gain/loss are also provided in Supplementary Information.

## DISCUSSION

In conclusion, we have predicted an anomalous light scattering property associated with the complex topological edge modes supported in a near-$\mathcal{PT}$-symmetric plasmonic system. Due to its special $\mathcal{PT}$-symmetric properties, such an anomalous scattering field is almost anti-symmetric when the system is excited by a symmetric plane wave. This resonantly scattered field has no contribution to the transmission and reflection parallel to the incident light, which means that the total forward extinction and feedback to the source is near zero. Since the resonant scattering is associated with the topological zero modes (which maps to a fixed frequency), the frequency at which the anomalous scattering occurs could be robust against weak perturbations. The result suggests an extreme situation where leakage of classical information is unnoticeable to the transmitter and the receiver. Although only plasmonic examples are given in this paper, the wave phenomenon could be general to other similar $\mathcal{PT}$-symmetric systems.

## METHODS

### Analytical evaluation of Zak phase.

We evaluate the Zak phase analytically. The Zak phases discussed in all figures are evaluated using Eq. 3. The derivation of Eq. 3 (in usual basis) is provided here. For reference, the Zak phase in bi-orthonormal basis are also provided in Supplementary Information. To derive Eq. (4), we consider the eigenvalues and the right eigenvectors of Eq. (1), which are $E_{k\pm} = \pm\sqrt{h_x^2 + h_y^2 - h_z^2}$ and

$$\mathbf{u}_\pm^R = \frac{1}{\sqrt{N_\pm}} \begin{bmatrix} h_x(k) - ih_y(k) \\ E_{k\pm} - ih_z(k) \end{bmatrix}, \quad (12)$$

where the normalizing factor is defined as $N_\pm := (h_x - ih_y)^*(h_x - ih_y) + (E_{k\pm} - ih_z)^*(E_{k\pm} - ih_z) = 2(h_x^2 + h_y^2)$, so that $(\mathbf{u}_\pm^R)^* \cdot \mathbf{u}_\pm^R = 1$. Although $(\mathbf{u}_\mp^R)^* \cdot \mathbf{u}_\pm^R \neq 0$, we still can evaluate the Zak phase of each band as those defined in Hermitian systems according to Eq. (3). We also provide another version of Zak phase defined in bi-orthonormal basis in Supplementary Information as a comparison.

Here, we parameterize the solution by putting $h_x = h\sin\theta\cos\phi$, $h_y = h\sin\theta\sin\phi$, and $h_z = h\cos\theta$, where $\theta$ and $\phi$ are azimuthal and polar angle of $\vec{h}(k)$, and $h := (h_x^2 + h_y^2 + h_z^2)^{1/2}$. As a result, the right eigenvectors in Eq. (12) are simplified as

$$\mathbf{u}_\pm^R = \frac{1}{\sqrt{2}} \begin{bmatrix} e^{-i\phi(k)} \\ \pm\sqrt{1 - \cot^2\theta(k)} - i\cot\theta(k) \end{bmatrix}. \quad (13)$$

Using Eq. (13), we can write the integrant in the integral of Eq. (3) as $\frac{1}{2}e^{i\phi(k)}\frac{d}{dk}e^{-i\phi(k)} + \frac{1}{2}[\pm\sqrt{1 - \cot^2\theta(k)} + i\cot\theta(k)] \times \frac{d}{dk}[\pm\sqrt{1 - \cot^2\theta(k)} - i\cot\theta(k)]$. Note that the condition $h_x^2 + h_y^2 > h_z^2$ implies $\sqrt{1 - \cot^2\theta(k)}$ is real for all $k$, and by defining $\xi(k) := \arg[\sqrt{1 - \cot^2\theta(k)} + i\cot\theta(k)]$, the second term can be simplified into $\frac{1}{2}e^{\pm i\xi(k)}\frac{d}{dk}e^{\mp i\xi(k)} = \mp\frac{i}{2}\frac{d\xi}{dk}$. As a result, the Zak phase in Eq. (3) becomes

$$\begin{aligned} \gamma_\pm &= \frac{1}{2}\int_{-\frac{\pi}{d}}^{\frac{\pi}{d}} dk \left[\frac{d}{dk}\phi(k) \pm \frac{d}{dk}\xi(k)\right] \\ &= \frac{1}{2}\left[\phi(k)|_{-\pi/d}^{\pi/d} \pm \xi(k)|_{-\pi/d}^{\pi/d}\right]. \end{aligned} \quad (14)$$

Since $\mathcal{C}$ is closed, $\phi(k)|_{-\pi/d}^{\pi/d}$ and $\xi(k)|_{-\pi/d}^{\pi/d}$ are integer multiples of $2\pi$. Furthermore, since $\text{Re}[e^{i\xi(k)}] = \sqrt{1 - \cot^2\theta(k)} > 0$ for all $k$, its argument is bounded $(-\pi/2 < \xi(k) < \pi/2)$, which means $\xi(k)|_{-\pi/d}^{\pi/d} = 0$. As a result, Eq. (14) gives Eq.(4).

### Coupled-dipole method.

The particle array is modeled as an array of point dipole scatters embedded in air for the sake of simplicity. By the next nearest neighbor approximation together with quasistatic approximation, the dipole moment $p_{n;\sigma}$ ($x$ component) induced on the nanoparticle $\sigma = A$ or $B$ in the $n$th unit cell follows the coupled dipole equation[23,33,44,45]:

$$\begin{cases} \alpha_A^{-1} p_{n;A} = \frac{2}{4\pi\epsilon_0}\left(\frac{1}{|d-s|^3}p_{n-1;B} + \frac{1}{|s|^3}p_{n;B}\right) + E_0 \\ \alpha_B^{-1} p_{n;B} = \frac{2}{4\pi\epsilon_0}\left(\frac{1}{|s|^3}p_{n;A} + \frac{1}{|d-s|^3}p_{n+1;A}\right) + E_0 \end{cases}, \quad (15)$$

where $n = 1, 2, ..., N$ and $E_0$ denotes the $x$-polarized external driving electric field.

Outer and inner radii of the coated nanoparticles are $b$ and $a$, and the inverse quasistatic polarizabilites are given by[46]

$$\alpha_A^{-1} = \frac{1}{4\pi\epsilon_0 b^3} \frac{c_1\epsilon_1 + c_2}{c_3\epsilon_1 + c_4} \quad \text{and} \quad \alpha_B^{-1} = \frac{1}{4\pi\epsilon_0 b^3} \frac{c_1^*\epsilon_1 + c_2^*}{c_3^*\epsilon_1 + c_4^*}, \quad (16)$$

in which $c_1 = (\epsilon_3 + 2) + 2(\epsilon_3 - 1)(a/b)^3$, $c_2 = 2(\epsilon_3 + 2)\epsilon_3 - 2(\epsilon_3 - 1)\epsilon_3(a/b)^3$, $c_3 = (\epsilon_3 - 1) + (2\epsilon_3 + 1)(a/b)^3$, and $c_4 = 2\epsilon_3(\epsilon_3 - 1) + \epsilon_3(2\epsilon_3 + 1)(a/b)^3$. Note that $c_i$ are constants that are independent of $\epsilon_1$.

## Formulation of eigenvalue problem for finite array.

The eigenvalue problem in the form similar to Bergman's representation is an important step to connect the actual problem to the topological band theory. Our formulation of eigenvalue problem is based on coupled-dipole method. It should be noted that obtaining resonant frequencies through this eigenvalue problem is much faster than searching complex non-trivial solutions directly using Eq. (15).

To do this, we group terms with $\epsilon_1$ in Eq. (15). The first and the second line are multiplied by $c_3\epsilon_1 + c_4$ and $c_3^*\epsilon_1 + c_4^*$ with respectively. By factoring out $\epsilon_1$, we have

$$\begin{cases} \epsilon_1 \left( \frac{c_3(b/d)^3}{|1-s/d|^3} p_{n-1;B} + c_1 p_{n;A} + \frac{c_3(b/d)^3}{|s/d|^3} p_{n;B} \right) = \frac{-c_4(b/d)^3}{|1-s/d|^3} p_{n-1;B} - c_2 p_{n;A} - \frac{c_4(b/d)^3}{|s/d|^3} p_{n;B} + (c_3\epsilon_1 + c_4)E_0 \\ \epsilon_1 \left( \frac{c_3^*(b/d)^3}{|s/d|^3} p_{n;A} + c_1^* p_{n;B} + \frac{c_3^*(b/d)^3}{|1-s/d|^3} p_{n+1;A} \right) = \frac{-c_4^*(b/d)^3}{|s/d|^3} p_{n;A} - c_2^* p_{n;B} - \frac{c_4^*(b/d)^3}{|1-s/d|^3} p_{n+1;A} + (c_3^*\epsilon_1 + c_4^*)E_0 \end{cases}$$

The above equations can be vectorized into the following matrix equation:

$$\epsilon_1 \mathbf{N}\mathbf{p} = \mathbf{M}\mathbf{p} + 4\pi\epsilon_0 b^3(\epsilon_1 \mathbf{C}_3 + \mathbf{C}_4)\mathbf{E}_0, \quad (17)$$

where $\mathbf{p} = (p_{1;A}, p_{1;B}, p_{2;A}, p_{2;B}, ..., p_{N;B})^{\mathrm{T}}$ and $\mathbf{E}_0 = (E_0, E_0, ..., E_0)^{\mathrm{T}}$. The $2N \times 2N$ matrixes are

$$\mathbf{C}_3 = \begin{bmatrix} c_3 & 0 & 0 & \cdots \\ 0 & c_3^* & 0 & \\ 0 & 0 & c_3 & \\ \vdots & & & \ddots \end{bmatrix}, \quad \mathbf{C}_4 = \begin{bmatrix} c_4 & 0 & 0 & \cdots \\ 0 & c_4^* & 0 & \\ 0 & 0 & c_4 & \\ \vdots & & & \ddots \end{bmatrix},$$

$$\mathbf{M} = \begin{bmatrix} -c_2 & \frac{-c_4(b/d)^3}{|s/d|^3} & 0 & 0 & \cdots \\ \frac{-c_4^*(b/d)^3}{|s/d|^3} & -c_2^* & \frac{-c_4^*(b/d)^3}{|1-s/d|^3} & 0 & \\ 0 & \frac{-c_4(b/d)^3}{|1-s/d|^3} & -c_2 & \frac{-c_4(b/d)^3}{|s/d|^3} & \\ 0 & 0 & \frac{-c_4^*(b/d)^3}{|s/d|^3} & -c_2^* & \\ \vdots & & & & \ddots \end{bmatrix}, \quad (18)$$

$$\mathbf{N} = \begin{bmatrix} c_1 & \frac{c_3(b/d)^3}{|s/d|^3} & 0 & 0 & \cdots \\ \frac{c_3^*(b/d)^3}{|s/d|^3} & c_1^* & \frac{c_3^*(b/d)^3}{|1-s/d|^3} & 0 & \\ 0 & \frac{c_3(b/d)^3}{|1-s/d|^3} & c_1 & \frac{c_3(b/d)^3}{|s/d|^3} & \\ 0 & 0 & \frac{c_3^*(b/d)^3}{|s/d|^3} & c_1^* & \\ \vdots & & & & \ddots \end{bmatrix}. \quad (19)$$

Multiplying Eq. (17) by $\mathbf{N}^{-1}$, we have

$$\mathbf{N}^{-1}\mathbf{M}\mathbf{p} = \epsilon_1 \mathbf{p} - 4\pi\epsilon_0 b^3 \mathbf{N}^{-1}(\epsilon_1 \mathbf{C}_3 + \mathbf{C}_4)\mathbf{E}_0. \quad (20)$$

Rearranging, it becomes

$$\mathbf{p} = \mathbf{S}(\epsilon_1)\mathbf{E}_0, \quad (21)$$

where $\mathbf{S}(\epsilon_1) = -4\pi\epsilon_0 b^3 (\mathbf{N}^{-1}\mathbf{M} - \epsilon_1 \mathbf{I}_{2N})^{-1}\mathbf{N}^{-1}(\epsilon_1 \mathbf{C}_3 + \mathbf{C}_4)$. This recovers Eq. (8).

To look for resonant modes, we set $\mathbf{E}_0 = 0$ in Eq. (20), which gives an eigenvalue problem,

$$\mathbf{N}^{-1}\mathbf{M}\mathbf{p} = \epsilon_1 \mathbf{p}. \quad (22)$$

The $2N$ possible eigenvalues $\epsilon_1$ are mapped to $\omega$ by Eq. (6), giving $2N$ resonant frequencies, see Fig. 2(b); while the mode patterns are the $2N$ eigenvectors, see Fig. 2(c).

## Formulation of eigenvalue problem for infinite array.

The formulation of the 2-band eigenvalue problem for an infinite array is again based on coupled-dipole method with the use of Bloch's theorem. We first put $E_0 = 0$ in Eq. (15). We write $p_{n;A} = p_{k;A}e^{iknd}$ and $p_{n;B} = p_{k;B}e^{iknd}$, and Eq. (15) becomes

$$\begin{cases} 4\pi\epsilon_0 b^3 \alpha_A^{-1} p_{k;A} = G_{k;AB} p_{k;B} \\ 4\pi\epsilon_0 b^3 \alpha_B^{-1} p_{k;B} = G_{-k;AB} p_{k;A} \end{cases} \quad (23)$$

where $G_{k;AB} = 2(b/d)^3 [e^{-ikd}/(1-s/d)^3 + 1/(s/d)^3]$. To obtain an eigenvalue problem with $\epsilon_1$ as the eigenvalue, we group terms with $\epsilon_1$. Eq. (23) is then vectorized to the matrix equation $\epsilon_1 \mathbf{N}_k \mathbf{p}_k = \mathbf{M}_k \mathbf{p}_k$, in which $\mathbf{p}_k = (p_{k;A}, p_{k;B})^{\mathrm{T}}$,

$$\mathbf{M}_k = \begin{bmatrix} -c_2 & c_4 G_{k;AB} \\ c_4^* G_{k;AB}^* & -c_2^* \end{bmatrix}, \quad (24)$$

and

$$\mathbf{N}_k = \begin{bmatrix} c_1 & -c_3 G_{k;AB} \\ -c_3^* G_{k;AB}^* & c_1^* \end{bmatrix}. \quad (25)$$

Multiplying the matrix equation by $\mathbf{N}_k^{-1}$, we recover the eigenvalue problem shown in Eq. (5):

$$\mathbf{A}_k \mathbf{p}_k = \epsilon_1 \mathbf{p}_k, \quad (26)$$

where

$$\begin{aligned}\mathbf{A}_k &:= \mathbf{N}_k^{-1} \mathbf{M}_k \\ &= \frac{1}{\Delta} \begin{bmatrix} c_3 c_4^* |G_{k;AB}|^2 - c_2 c_1^* & (c_4 c_1^* - c_3 c_2^*) G_{k;AB} \\ (c_1 c_4^* - c_2 c_3^*) G_{k;AB}^* & c_4 c_3^* |G_{k;AB}|^2 - c_1 c_2^* \end{bmatrix}\end{aligned}$$
(27)

and $\Delta = c_1 c_1^* - c_3 c_3^* |G_{k;AB}|^2$ (which is always real). As $G_{k+2\pi/d;AB} = G_{k;AB}$, we can write $\mathbf{A}_k = f_k \mathbf{I}_2 + \mathbf{H}_k$, where $f_k = \text{Re}[(c_3 c_4^* |G_{k;AB}|^2 - c_2 c_1^*)/\Delta]$, $h_z(k) = \text{Im}[(c_3 c_4^* |G_{k;AB}|^2 - c_2 c_1^*)/\Delta]$, $h_x(k) = \text{Re}\{[(c_4 c_1^* - c_3 c_2^*) G_{k;AB}]/\Delta\}$, $h_y(k) = \text{Im}\{[(c_1 c_4^* - c_2 c_3^*) G_{k;AB}^*]/\Delta\}$, and $\mathbf{I}_2$ is a $2 \times 2$ identity matrix.

**Time reversal operator $\mathcal{T}$.**

We define the time-reversal operator $\mathcal{T}$ for the analysis of $\mathcal{PT}$ symmetry in the basis of coupled-dipole method. Suppose the time varying dipole moment on the $i$th nanoparticle has the general form $p_i(t) = \text{Re} \int_{-\infty}^{\infty} p_i(\omega) e^{-i\omega t} d\omega$, where $\omega$ is the angular frequency and $p_i(\omega)$ is the Fourier component. Defining the time reverse operator by $\mathcal{T} p_i(t) = p_i(-t)$, we have

$$\begin{aligned}\mathcal{T} p_i(t) &= \text{Re} \int_{-\infty}^{\infty} p_i(\omega) e^{i\omega t} d\omega = \text{Re}[\int_{-\infty}^{\infty} p_i(\omega)^* e^{-i\omega t} d\omega]^* \\ &= \text{Re}[\int_{-\infty}^{\infty} p_i(\omega)^* e^{-i\omega t} d\omega].\end{aligned}$$
(28)

From the above, $\mathcal{T}$ is thus turning the Fourier component into its complex conjugate, which means $\mathcal{T} p_i = p_i^*$.

For the case that the plasmonic cores are lossless, time reverse operation turns the energy gaining dielectric into energy losing dielectric and vice versa, which means the time reversed particle array is effectively the same as the particle array obtained by swapping the positions of nanoparticles $A$ and $B$.

**Analysis of parity-time $\mathcal{PT}$ symmetry of matrix.**

The analysis of $\mathcal{PT}$ symmetry is again in the basis of coupled-dipole method. Here we would like to show that $\mathbf{S}(\epsilon_1)$ in Eq. (8) and $\mathcal{PT}$ commute when $\epsilon_1 \in \mathbb{R}$.

Firstly, if a general $2N \times 2N$ matrix $\mathbf{A}$ commutes with $\mathcal{PT}$, then $\mathbf{A}^{-1}$ commutes with $\mathcal{PT}$ also. To show this, we consider the fact that $\mathbf{AA}^{-1} = \mathbf{I}_{2N}$, we have $(\mathcal{PT})\mathbf{A}(\mathcal{PT})^{-1}(\mathcal{PT})\mathbf{A}^{-1}(\mathcal{PT})^{-1} = \mathbf{I}_{2N}$. Since $\mathbf{A}$ commutes with $\mathcal{PT}$ [i.e., $(\mathcal{PT})\mathbf{A}(\mathcal{PT})^{-1} = \mathbf{A}$], we have $\mathbf{A}(\mathcal{PT})\mathbf{A}^{-1}(\mathcal{PT})^{-1} = \mathbf{I}_{2N}$. Further multiplying $\mathbf{A}^{-1}$ from the left, we have

$$(\mathcal{PT})\mathbf{A}^{-1}(\mathcal{PT})^{-1} = \mathbf{A}^{-1}. \quad (29)$$

Secondly, if matrixes $\mathbf{A}$ and $\mathbf{B}$ commute with $\mathcal{PT}$, then their combination, $\mathbf{AB}$, also commutes with $\mathcal{PT}$:

$$\begin{aligned}(\mathcal{PT})\mathbf{AB}(\mathcal{PT})^{-1} &= (\mathcal{PT})\mathbf{A}(\mathcal{PT})^{-1}(\mathcal{PT})\mathbf{B}(\mathcal{PT})^{-1} \\ &= \mathbf{AB}.\end{aligned}$$
(30)

It is obvious to see from their definitions, $\mathbf{N}$, $\mathbf{M}$, $\mathbf{C}_3$, and $\mathbf{C}_4$ commute with $\mathcal{PT}$. Thus, by the rules stated in Eqs. (29) and (30), we have $(\mathcal{PT})\mathbf{N}^{-1}\mathbf{M}(\mathcal{PT})^{-1} = \mathbf{N}^{-1}\mathbf{M}$. Extending the consideration gives

$$\begin{aligned}&(\mathcal{PT})(\mathbf{N}^{-1}\mathbf{M} - \epsilon_1 \mathbf{I}_{2N})(\mathcal{PT})^{-1} \\ &= (\mathcal{PT})\mathbf{N}^{-1}\mathbf{M}(\mathcal{PT})^{-1} - (\mathcal{PT})\epsilon_1 \mathbf{I}_{2N}(\mathcal{PT})^{-1} \quad (31) \\ &= \mathbf{N}^{-1}\mathbf{M} - (\mathcal{PT})\epsilon_1(\mathcal{PT})^{-1} \mathbf{I}_{2N}.\end{aligned}$$

We see if $\epsilon_1$ is real, we have $(\mathcal{PT})\epsilon_1(\mathcal{PT})^{-1} = \epsilon_1$, and therefore $\mathbf{N}^{-1}\mathbf{M} - \epsilon_1 \mathbf{I}_{2N}$ commutes with $\mathcal{PT}$. Similarly, $\epsilon_1 \mathbf{C}_3 + \mathbf{C}_4$ commutes with $\mathcal{PT}$ if $\epsilon_1$ is real. As a result, again by Eqs. (29) and (30), recalling the definition of $\mathbf{S}(\epsilon_1)$ below Eq. (21), $\mathbf{S}(\epsilon_1)$ commutes with $\mathcal{PT}$ if $\epsilon_1$ is real.

**Analytical solutions of topological edge modes.**

The analytical solutions of the topological edge modes are based on coupled-dipole method in quasi-static approximation. For simplicity, we consider a semi-infinite plasmonic particle array with only one edge on the left. The protected edge modes are required to satisfy both the boundary condition and the dispersion relation with complex wave vectors $kd = \pm\pi + i\kappa d$[23,48], where $\kappa \in \mathbb{R}$.

By assuming the dipole moments are in the Bloch's form with complex wave vectors, we have

$$p_{n;\sigma} = p_{k;\sigma} e^{iknd} = (-1)^n p_{k;\sigma} e^{-n\kappa d}, \quad (32)$$

where $\sigma = A$ or $B$. The dipole moments in the first two unit cells have to follow the coupled dipole equation at the boundary:

$$\begin{cases} \alpha_A^{-1} p_{1;A} = \dfrac{2}{4\pi\epsilon_0 s^3} p_{1;B} \\ \alpha_B^{-1} p_{1;B} = \dfrac{2}{4\pi\epsilon_0 s^3} p_{1;A} + \dfrac{2}{4\pi\epsilon_0 (d-s)^3} p_{2;A}, \end{cases} \quad (33)$$

which is simply obtained by putting $p_{0;\sigma} = 0$ in Eq. (15). Substituting $p_{2;A} = -p_{1;A} e^{\kappa d}$ to Eq. (33), and eliminating the fraction $p_{1;B}/p_{1;A}$, we have

$$\alpha_A^{-1} \alpha_B^{-1} = \left(\frac{2}{4\pi\epsilon_0}\right)^2 \left(\frac{1}{s^6} - \frac{e^{-\kappa d}}{s^3(d-s)^3}\right), \quad (34)$$

which is a condition obtained by considering the boundary. On the other hand, the existence of the nontrivial solution $p_{k;\sigma}$ in Eq. (23) implies $\alpha_A^{-1} \alpha_B^{-1} = G_{k;AB} G_{-k;AB}$.

When $k$ is real, it gives the dispersion relation; when $kd = \pm\pi + \kappa d$, it becomes

$$\alpha_A^{-1}\alpha_B^{-1} = \left(\frac{2}{4\pi\epsilon_0}\right)^2 \left(\frac{1}{(d-s)^6} + \frac{1}{s^6} - \frac{e^{\kappa d} + e^{-\kappa d}}{s^3(d-s)^3}\right), \quad (35)$$

which is another condition for the edge state.

Eliminating $\alpha_A^{-1}\alpha_B^{-1}$ by Eqs. (34) and (35), we obtain the decaying factor

$$e^{-\kappa d} = (d-s)^3/s^3. \quad (36)$$

As only decaying dipole moments are physical ($\kappa \geq 0$), there is no solution for $\kappa$ unless $s > 0.5d$, which is the case with non-trivial Zak phase. By putting Eq. (36) back into Eq. (34), we have either $\alpha_A^{-1} = 0$ or $\alpha_B^{-1} = 0$. However, since $\alpha_B^{-1} = 0$ leads to the trivial solution $p_{1;A} = p_{1;B} = 0$ in Eq. (33), we have $\alpha_A^{-1} = 0$. This implies $\epsilon_1 = -c_2/c_1$, $p_{1;A} = 1$, and $p_{1;B} = 0$. Putting back to the Drude model Eq. (6), we have

$$\omega/\omega_p = \frac{1}{2}\left(\sqrt{\frac{4}{1+c_2/c_1} - \frac{1}{\tau^2}} - \frac{1}{\tau}i\right), \quad (37)$$

which is the left edge state resonant frequency. Solution for the particle array with a right edge can be obtained similarly. In this case, $e^{-\kappa d} = s^3/(d-s)^3$ with $\kappa < 0$, and $\omega/\omega_p = \frac{1}{2}\sqrt{4/(1+c_2^*/c_1^*) - 1/\tau^2} - i/\tau$.

## ACKNOWLEDGEMENTS

This work was supported by the Hong Kong Research Grant Council through grant no. 509813, 15300315, and AoE/P-02/12. We thank Jack Ng and C. T. Chan for useful discussions.

## AUTHOR CONTRIBUTIONS

All authors discuss the work thoroughly and extensively. C.W.L. performed the analytical calculations. K.H.C. and T.C.M. contributed to the simulation and verification of the anomalous scattering phenomenon. Z.Q.Z contributed to the approaches of defining Zak phase. K.H.F. conceived the idea and oversaw the progress.

## ADDITIONAL INFORMATION

**Supplementary Information** accompanies this paper.
**Competing financial interests:** The authors declare no competing financial interests.

# Anomalous Light Scattering by Topological $\mathcal{PT}$-symmetric Particle Arrays: Supplementary Information

C. W. Ling,[1] Ka Hei Choi,[1] T. C. Mok,[1] Z. Q. Zhang,[2] and Kin Hung Fung[1, *]

[1]*Department of Applied Physics, The Hong Kong Polytechnic University, Hong Kong, China*
[2]*Department of Physics, The Hong Kong University of Science and Technology, Hong Kong, China*

[*]khfung@polyu.edu.hk

## A. ZAK PHASE AND BAND DISPERSION

In Fig. A.1(a), we show how the Zak phase $\gamma$ (as defined in Eq. (3)) changes as the non-Hermiticity $\text{Im}(\epsilon_3)$ increases. When $\text{Im}(\epsilon_3) > 0.16$, BZ contains broken $\mathcal{PT}$-symmetric phase, and thus $\gamma$ is not quantized. When $\text{Im}(\epsilon_3) < 0.16$, bands with $s = 0.6d$ are non-trivial, which gives the protected edge modes (integral paths $\vec{h}(k)$ are shown in Fig. B.1. Bulk band dispersions with $\text{Im}(\epsilon_3) = 0.025$ and $0.25$ are demonstrated in Fig. A.1(b) for reference, and in which only (b)(ii) contains the exceptional points. Note that bulk dispersions for an array with $s = 0.4d$ and $s = 0.6d$ are the same, as the two infinite arrays only differ in a geometrical shift with $d/2$.

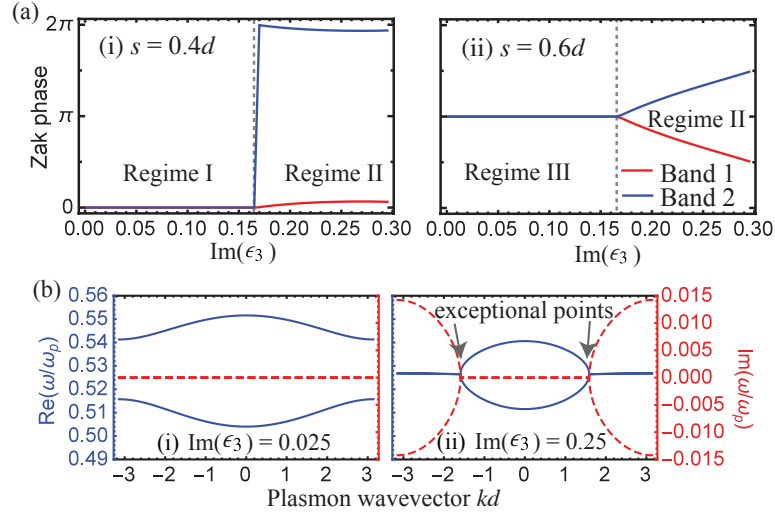

FIG. A.1. (Color online) (a) Zak phases $\gamma$ with (i) $s = 0.4d$ and (ii) $s = 0.6d$. $\text{Re}(\epsilon_3)=1.5$. When BZ contains broken $\mathcal{PT}$-symmetric phase (regime II), $\gamma$ is not quantized. Otherwise, it is either classified as trivial (regime I, $\gamma = 0$) or non-trivial (regime II, $\gamma = \pi$). (b) Corresponding bulk dispersion relation for $s = 0.4d$ or $s = 0.6d$ when (i) $\text{Im}(\epsilon_3) = 0.025$ and (ii) $\text{Im}(\epsilon_3) = 0.25$.

## B. EXACT $\vec{h}$-SPACE DIAGRAM FOR PLASMONIC DIMER ARRAYS

A closed loop is formed by $\vec{h}(k)$ when $kd$ changes from $-\pi$ to $\pi$, as shown in Fig. B.1. Here, $\mathbf{H}_k$ has the form of Eq. (12). Since the additional term $f_k \mathbf{I}_2$ does not alter the eigenvectors, $\mathbf{A}_k$ share the same eigenvectors and Zak phase $\gamma$ with $\mathbf{H}_k$. The two eigenvalues are mapped to $\omega$ via Eq. (6), which gives the dispersion relation in Fig. A.1(b).

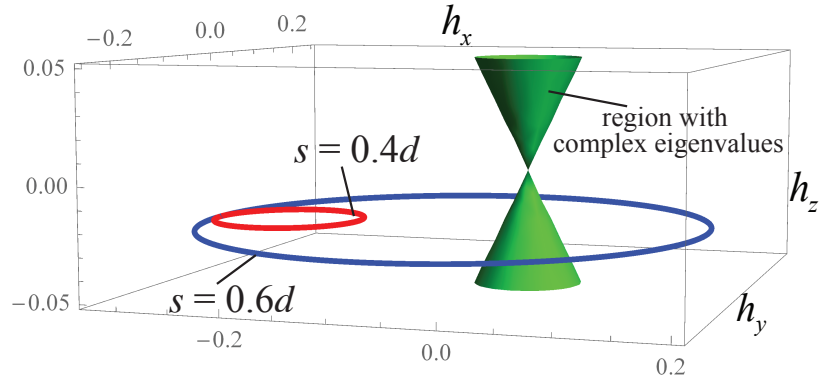

FIG. B.1. (Color online) Numerical integral path $\vec{h}(k)$ of an infinite plasmonic particle array. Parameters are $a = 0.125d$, $b = 0.175d$, $1/\tau = 0$, $\epsilon_3 = 1.5+0.025i$. The region within the kissing cones is where eigenvalues are complex, which corresponds to the broken $\mathcal{PT}$-symmetric phase.

## C. ZAK PHASE WITH BI-ORTHONORMAL BASIS

If bi-orthonormal bases are used to define $\gamma$ instead of the usual definition, $\gamma$ will be in general a complex number. However, its real part gives back the value obtained by Eq. (3). Here, we provide the evaluation of Zak phase based on the bi-orthornormal basis.

Recalling $\mathbf{H}_k$ in Eq. (1) is non-Hermitian, therefore, left eigenvectors have to be used to form a bi-orthonormal basis. The left eigenvector $\mathbf{u}^L$ satisfies the eigenvalue problem $\mathbf{H}_k^T \mathbf{u}^L = E_k \mathbf{u}^L$ [1–4], whose eigenvalues are $E_{k\pm} = \pm(h_\parallel^2 - h_z^2)^{1/2}$, where $h_\parallel = (h_x^2 + h_y^2)^{1/2}$. The right and left eigenvectors of $\mathbf{H}_k$ are

$$\mathbf{u}_\pm^R = \frac{1}{M_{k\pm}^{1/2}} \begin{bmatrix} h_x(k) - ih_y(k) \\ E_{k\pm} - ih_z(k) \end{bmatrix} \tag{C.1a}$$

and

$$\mathbf{u}_\pm^L = \frac{1}{M_{k\pm}^{1/2}} \begin{bmatrix} h_x(k) + ih_y(k) \\ E_{k\pm} - ih_z(k) \end{bmatrix}, \tag{C.1b}$$

where the normalizing factor $M_{k\pm} := (h_x - ih_y)(h_x + ih_y) + (E_{k\pm} - ih_z)^2 = h_\parallel^2 + (E_{k\pm} - ih_z)^2 = 2h_\parallel^2 - 2h_z^2 \pm 2ih_z\sqrt{h_\parallel^2 - h_z^2}$. We note that $M_{k\pm}$ is chosen such that the biorthnormal conditions $\mathbf{u}_\pm^L \cdot \mathbf{u}_\pm^R = 1$ and $\mathbf{u}_\mp^L \cdot \mathbf{u}_\pm^R = 0$ are satisfied.

The Zak phase defined by using biorthonormal basis is [1, 2, 4]

$$\gamma_\pm = i \int_{-\frac{\pi}{d}}^{\frac{\pi}{d}} dk \left( \mathbf{u}_\pm^L \cdot \frac{d}{dk} \mathbf{u}_\pm^R \right), \tag{C.2}$$

which is similar to Eq. (3). In this case, $\gamma_\pm$ will be a complex number even the entire bulk dispersion is in the unbroken $\mathcal{PT}$-symmetric phase, and its real part is the same as that in Eq. (4). To show this, we again restrict $E_{k\pm}$ are real. We put $h_x(k) + ih_y(k) = h_\parallel(k) e^{i\phi(k)}$. Using the product rule, Eq. (C.2) becomes

$$\begin{aligned}\gamma_\pm =& i \int_{-\frac{\pi}{d}}^{\frac{\pi}{d}} dk \left( \frac{h_\parallel e^{i\phi}}{M_{k\pm}} \frac{d}{dk} h_\parallel e^{-i\phi} \right. \\ & + \frac{h_\parallel^2}{M_{k\pm}^{1/2}} \frac{d}{dk} \frac{1}{M_{k\pm}^{1/2}} + \frac{E_{k\pm} - ih_z}{M_{k\pm}} \frac{d}{dk}(E_{k\pm} - ih_z) \\ & \left. + \frac{(E_{k\pm} - ih_z)^2}{M_{k\pm}^{1/2}} \frac{d}{dk} \frac{1}{M_{k\pm}^{1/2}} \right). \end{aligned} \tag{C.3}$$

The first term in the integrant gives

$$\begin{aligned} \frac{h_\parallel e^{i\phi}}{M_{k\pm}} \frac{d}{dk} h_\parallel e^{-i\phi} &= \frac{-ih_\parallel^2}{M_{k\pm}} \frac{d}{dk}\phi + \frac{h_\parallel}{M_{k\pm}} \frac{d}{dk} h_\parallel \\ &= \frac{-ih_\parallel^2}{M_{k\pm}} \frac{d}{dk}\phi + \frac{1}{2M_{k\pm}} \frac{d}{dk} h_\parallel^2. \end{aligned} \tag{C.4a}$$

The second and the forth terms give

$$\begin{aligned} & \frac{h_\parallel^2}{M_{k\pm}^{1/2}} \frac{d}{dk} \frac{1}{M_{k\pm}^{1/2}} + \frac{(E_{k\pm} - ih_z)^2}{M_{k\pm}^{1/2}} \frac{d}{dk} \frac{1}{M_{k\pm}^{1/2}} \\ &= \frac{h_\parallel^2 + (E_{k\pm} - ih_z)^2}{M_{k\pm}^{1/2}} \frac{d}{dk} \frac{1}{M_{k\pm}^{1/2}} \\ &= M_{k\pm}^{1/2} \frac{d}{dk} \frac{1}{M_{k\pm}^{1/2}} = -\frac{1}{2M_{k\pm}} \frac{d}{dk} M_{k\pm}, \end{aligned} \tag{C.4b}$$

in which we used the definition of $M_{k\pm}$ [below Eq. (C.1)]. The third term gives

$$\frac{E_{k\pm} - ih_z}{M_{k\pm}} \frac{d}{dk}(E_{k\pm} - ih_z) = \frac{1}{2M_{k\pm}} \frac{d}{dk}(E_{k\pm} - ih_z)^2. \tag{C.4c}$$

Substituting Eqs. (C.4) into Eq. (C.3), we have

$$
\begin{aligned}
\gamma_\pm =& i\int_{-\frac{\pi}{d}}^{\frac{\pi}{d}} dk \left( \frac{-ih_\parallel^2}{M_{k\pm}} \frac{d}{dk}\phi + \frac{1}{2M_{k\pm}} \frac{d}{dk} h_\parallel^2 \right. \\
& \left. -\frac{1}{2M_{k\pm}} \frac{d}{dk} M_{k\pm} + \frac{1}{2M_{k\pm}} \frac{d}{dk} (E_{k\pm} - ih_z)^2 \right) \\
=& i\int_{-\frac{\pi}{d}}^{\frac{\pi}{d}} dk \left( \frac{-ih_\parallel^2}{M_{k\pm}} \frac{d}{dk}\phi + \frac{1}{2M_{k\pm}} \frac{d}{dk} M_{k\pm} \right. \\
& \left. -\frac{1}{2M_{k\pm}} \frac{d}{dk} M_{k\pm} \right) = \int_{-\frac{\pi}{d}}^{\frac{\pi}{d}} dk \frac{h_\parallel^2}{M_{k\pm}} \frac{d}{dk}\phi.
\end{aligned}
\quad (C.5)
$$

Noticing that as long as $E_{k\pm}$ are real, $(h_\parallel^2 - h_z^2)^{1/2}$ are also real. Then, by rationalizing the fraction in the last line of Eq. (C.5), we have

$$
\begin{aligned}
\frac{h_\parallel^2}{M_{k\pm}} &= \frac{1}{2} \frac{h_\parallel^2}{h_\parallel^2 - h_z^2 \pm ih_\parallel^2 h_z (h_\parallel^2 - h_z^2)^{1/2}} \\
&= \frac{1}{2} \frac{h_\parallel^2 \mp ih_\parallel^2 h_z / (h_\parallel^2 - h_z^2)^{1/2}}{h_\parallel^2} \\
&= \frac{1}{2} \mp i\frac{h_z}{2(h_\parallel^2 - h_z^2)^{1/2}}.
\end{aligned}
\quad (C.6)
$$

Finally, by putting Eq. (C.6) into Eq. (C.5), we have

$$
\begin{aligned}
\gamma_\pm &= \int_{-\pi/d}^{\pi/d} \frac{d\phi}{dk} dk \mp i\int_{-\pi/d}^{\pi/d} \frac{h_z}{2(h_\parallel^2 - h_z^2)^{1/2}} dk \\
&= w\pi \mp i\int_{-\pi/d}^{\pi/d} \frac{h_z}{2(h_\parallel^2 - h_z^2)^{1/2}} dk,
\end{aligned}
\quad (C.7)
$$

where $w$ is the winding number of $\vec{h}(k)$ about the $h_z$ axis. Eq. (C.7) shows that $\gamma$ is in general a complex number, unless $\vec{h}(k)$ has some other symmetries so that the integral vanishes.

## D.  FIELD PATTERNS FOR A NORMAL ARRAY ($\epsilon_3 = 1.5$) BY MST

For comparison with Fig. 5(b), here we show the electric field pattern of a normal array with $\epsilon_3 = 1.5$ in Fig. D.1. Since the normal array $\mathcal{P}$ symmetry, the response of the array should be symmetric, which gives an anti-symmetric pattern for the $E_y$ field component. The two figures (Figs. 5 and D.1) verified the strong antisymmetry response of the non-Hermitian particle array, which does not contribute to forward and backward scattering.

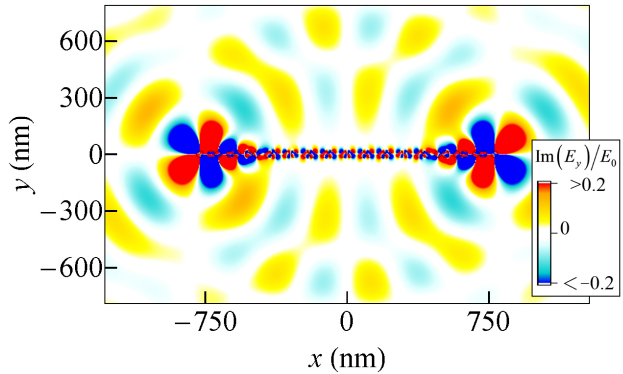

FIG. D.1. (Color online) Electric field pattern of the $E_y$ field component at edge mode frequency ($\omega = 0.5188\omega_p$) of the normal array (with $\epsilon_3 = 1.5$ and $s = 0.6d$).

17